\documentclass[conference]{IEEEtran}
\IEEEoverridecommandlockouts
\usepackage{cite}
\usepackage{amsmath,amssymb,amsfonts}
\usepackage{algorithmic}
\usepackage{graphicx}
\usepackage{url}
\usepackage{psfrag}
\usepackage{booktabs} 
\usepackage{multirow}
\usepackage{epsfig}
\usepackage{tabularx}
\usepackage{textcomp}
\usepackage{xcolor}
\def\BibTeX{{\rm B\kern-.05em{\sc i\kern-.025em b}\kern-.08em
    T\kern-.1667em\lower.7ex\hbox{E}\kern-.125emX}}

\newcommand{\secref}[1]{Section~\ref{#1}}

\newcommand{\figref}[1]{Fig.~\ref{#1}}
\newcommand{\tabref}[1]{Tab.~\ref{#1}}

\begin{document}

\title{LoopLynx: A Scalable Dataflow Architecture for Efficient LLM Inference\\
% {\footnotesize \textsuperscript{*}Note: Sub-titles are not captured in Xplore and
% should not be used}
\thanks{\textsuperscript{*}Corresponding author: Gang Chen, Email: cheng83@mail.sysu.edu.cn}
}

\author{\IEEEauthorblockN{Jianing Zheng, Gang Chen\textsuperscript{*}}
\IEEEauthorblockA{Sun Yat-sen University\\
% \textit{name of organization (of Aff.)}\\
% Guangzhou, China
% email address or ORCID
}
% \and
% \IEEEauthorblockN{2\textsuperscript{nd} Given Name Surname}
% \IEEEauthorblockA{\textit{dept. name of organization (of Aff.)} \\
% \textit{name of organization (of Aff.)}\\
% City, Country \\
% email address or ORCID}
% \and
% \IEEEauthorblockN{3\textsuperscript{rd} Given Name Surname}
% \IEEEauthorblockA{\textit{dept. name of organization (of Aff.)} \\
% \textit{name of organization (of Aff.)}\\
% City, Country \\
% email address or ORCID}
% \and
% \IEEEauthorblockN{4\textsuperscript{th} Given Name Surname}
% \IEEEauthorblockA{\textit{dept. name of organization (of Aff.)} \\
% \textit{name of organization (of Aff.)}\\
% City, Country \\
% email address or ORCID}
% \and
% \IEEEauthorblockN{5\textsuperscript{th} Given Name Surname}
% \IEEEauthorblockA{\textit{dept. name of organization (of Aff.)} \\
% \textit{name of organization (of Aff.)}\\
% City, Country \\
% email address or ORCID}
% \and
% \IEEEauthorblockN{6\textsuperscript{th} Given Name Surname}
% \IEEEauthorblockA{\textit{dept. name of organization (of Aff.)} \\
% \textit{name of organization (of Aff.)}\\
% City, Country \\
% email address or ORCID}
}

\maketitle

\begin{abstract}
In this paper, we propose LoopLynx, a scalable dataflow architecture
for efficient LLM inference that optimizes FPGA usage through a hybrid spatial-temporal design. The design of LoopLynx incorporates a hybrid temporal-spatial architecture, where computationally intensive operators are implemented as large dataflow kernels. This achieves high throughput similar to spatial architecture, and organizing and reusing these kernels in a temporal way together enhances FPGA peak performance. 
Furthermore, to overcome the resource limitations of a single device, we provide a multi-FPGA distributed architecture that overlaps and hides all data transfers so that the distributed accelerators are fully utilized. By doing so, LoopLynx can be effectively scaled to multiple devices to further explore model parallelism for large-scale LLM inference. Evaluation of GPT-2 model demonstrates that LoopLynx can achieve comparable performance to state-of-the-art single FPGA-based accelerations. In addition, compared to Nvidia A100, our accelerator with a dual-FPGA configuration delivers a 2.52x speed-up in inference latency while consuming only 48.1\% of the energy.
\end{abstract}

% \begin{IEEEkeywords}
% component, formatting, style, styling, insert
% \end{IEEEkeywords}

\section{Introduction}
Over recent years, we have witnessed a remarkable technological revolution in large language models (LLMs). Due to their powerful capabilities in natural language processing tasks and beyond, LLMs have transformed the way we interact with AI models. They have introduced new efficiencies and capabilities in many exciting downstream applications, such as chatbots~\cite{chatbot} and code generation.
The transformative power of LLMs stems from breakthroughs in the auto-regressive generation model. As illustrated in \figref{fig:llm-inference}, the auto-regressive process of LLM inference involves two stages: the prefill stage which processes the input prompt, and the decode stage which performs auto-regressive token generation. In the prefill stage, all tokens in the input sequence can be processed in parallel, effectively saturating GPU computing. In contrast, the decode stage processes only a single token at a time per request, limiting GPU to fully leverage its parallel capabilities.
To address the aforementioned challenges, FPGAs have been considered a promising solution for efficient LLM inference. Recent research on FPGA-based LLM accelerators \cite{dfx,FlightLLM,FRTRANS,NPE,Spatial} have demonstrated that a single FPGA can be competitive with GPUs in terms of inference latency and energy efficiency. According to~\cite{Spatial}, these accelerators can be classified into temporal and spatial architectures. Temporal architectures~\cite{dfx,FlightLLM,FRTRANS,NPE} 
reuse process engines (PEs) to execute instructions and employ overlay approaches to achieve efficient bitstream reuse across multiple models. However, such temporal architectures~\cite{dfx,FlightLLM,FRTRANS,NPE} require frequent off-chip memory access, resulting in the high cost of inference latency and energy consumption. In contrast, spatial architectures~\cite{Spatial} instantiate multiple PEs operating simultaneously in a dataflow mode,  substantially reducing off-chip memory accesses. However, the parallel processing capabilities of such dataflow architectures are largely underutilized in the decoding phase due to the sequential processing pattern. Moreover, the limited computational resources of a single FPGA make it difficult to achieve optimal end-to-end inference performance.  
To address the challenges, we propose LoopLynx, a scalable dataflow architecture that optimizes FPGA usage through a hybrid spatial-temporal design. LoopLynx combines the flexibility of temporal architecture for scheduling different PEs with the high throughput of spatial architecture. It implements computationally intensive operators as large dataflow kernels and uses a state machine to schedule and reuse these kernels, enhancing FPGA performance by increasing peak area utilization. In addition, we scale this hybrid architecture to multiple accelerator nodes which is possible to distribute to multi-FPGA platforms to explore the potential for parallel inference of large-scale LLM models. To achieve this, we distribute large-scale matrix operations across multiple nodes and interconnect them using a ring network, where the network synchronization overhead is effectively concealed within the dataflow design. Therefore, LoopLynx can achieve high scalability with minimal performance loss. Our main contributions are summarized as:
\begin{itemize}
	\item We propose LoopLynx, an efficient FPGA-based accelerator that uses hybrid architecture to compromise the flexibility and the throughput between temporal and spatial architecture.
	\item We explore the feasibility of implementing LoopLynx on a scalable multi-FPGA platform. 
\end{itemize}
We tested a GPT-2 model and compared its performance to the Nvidia A100 under the same quantization strategy. Our single-FPGA setup (with two accelerator nodes) achieves an average 1.67x speed-up over the Nvidia A100 while consuming only 37.3\% of its energy. When scaling to a dual-FPGA configuration through simulated network, we achieve a 2.52x speed-up, while consuming 48.1\% of the A100's energy. Compared to the state-of-the-art temporal architecture~\cite{dfx} and spatial architecture~\cite{Spatial}, our LoopLynx architecture with dual-FPGA configuration achieves 2.11x and 1.64x improvement in token generation latency, respectively.

\section{Related Work}
Previous work focused on accelerating the bidirectional attention model like BERT~\cite{TRAC,FQBert}, but this approach cannot support or fully utilize computational resources for auto-regressive inference of LLMs. 
To achieve the flexibility, most of LLM accelerators are designed as temporal architectures~\cite{NPE,FRTRANS,BalanceCompress,Codesign,ColomnPrune,EfficientMappint,EfficientTrans,dfx} which manage functional operations through an instruction set. However, such a sequential pattern of instruction execution prevents the FPGA's programmed areas from overlapping during the inference. FlightLLM~\cite{FlightLLM} also utilizes an instruction set architecture and accelerates larger models like LLaMA~\cite{llama} through model compression and sparse DSP chains. The work in~\cite{Spatial} provides a spatial architecture that connects multiple systolic arrays and other operators in a dataflow manner, activating all operators to establish a task-level pipeline to improve throughput during the prefill stage. However, the sequential processing patterns in the decoding stage of LLMs prevent continuous pipeline formation and thus hinder overall throughput. 
Compared to existing studies, we propose a scalable dataflow architecture for efficient LLM inference that optimizes FPGA usage through a hybrid spatial-temporal design. Our design combines the advantages of both spatial and temporal architectures~\cite{Spatial,dfx} to fully explore kernel parallelism for large-scale LLM inference.
\section{Background}
\begin{figure}[]
	\centering
	\includegraphics[width=\linewidth]{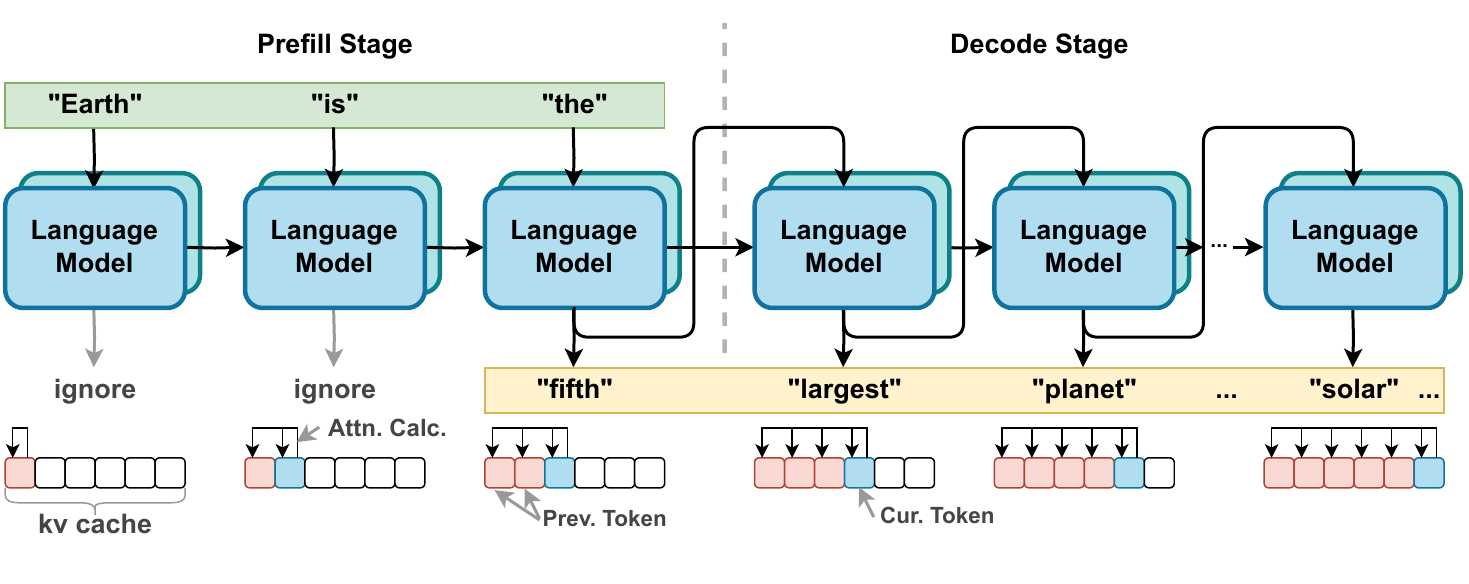}
	\vspace{-1.5\baselineskip}
	% \captionsetup{font=bf}
	\caption{Illustration of 2-stages of LLM inference}
	\vspace{-1.5\baselineskip}
	\label{fig:llm-inference}
\end{figure}
The inference process of auto-regressive LLMs consists of two stages: the prefill stage and the decode stage. During the prefill stage, the LLM processes user input prompts to fill the KV cache. Embedded prompts are passed through cascading transformer blocks~\cite{transformer}, with each block generating $K$ and $V$ matrices stored in their respective KV cache. During the decoding stage, the accumulated KV cache avoids repeatedly concatenating model outputs with prompts and recalculating previous tokens. Instead, it only requires computation for the newly generated token to obtain $Q$, $K$, and $V$ vector, and then compute attention with the cached KV values of previous tokens, thus reducing repetitive computations in the LLM generation stage.
We use \figref{fig:llm-inference} to better demonstrate the entire LLM working process. During the prefill stage, when given the prompt "Earth is the," the LLM reads each input token and calculates its attention with preceding tokens while excluding the output from the final transformer block. The output from this stage is not used because its primary function is to prepare the KV cache, which is needed for the subsequent decoding stage. After completing the last prefill inference, the generated token is now used as input for the decode stage. The model continues generating the sequence in an auto-regressive manner, producing tokens such as "fifth", "largest", "planet", "in", "the" and "universe" and it finishes upon encountering the end-of-sequence marker (EOS).

\begin{figure}[]
	\centering
	\includegraphics[width=1.0\linewidth]{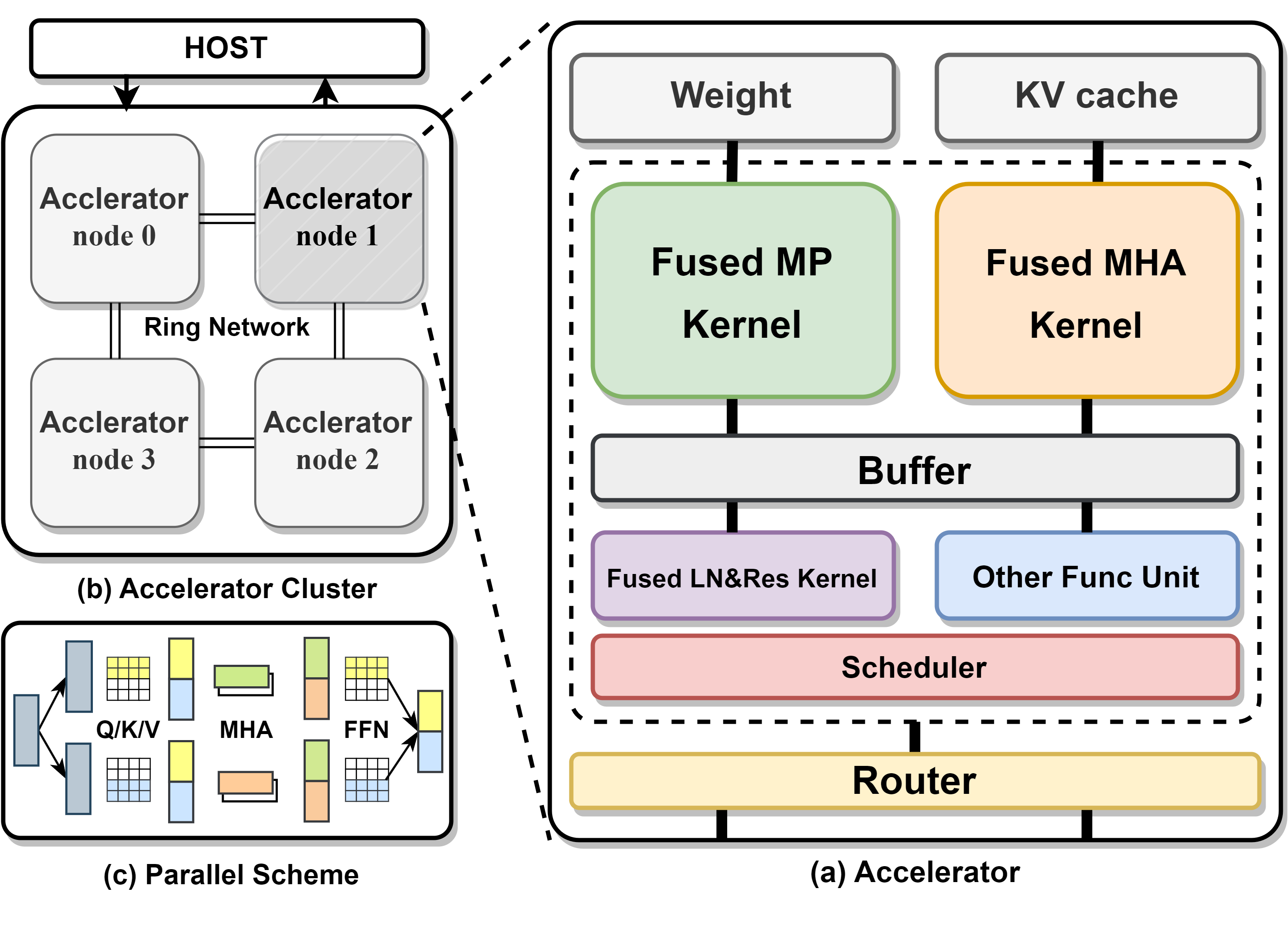}
	% \captionsetup{font=bf}
	\caption{The overall architecture of LoopLynx. Accelerators are connected via a ring network and operating under model parallel scheme.}
	\vspace{-1.5\baselineskip}
	\label{fig:architecture_overview}
\end{figure}
\begin{figure*}[h!]
	\centering
	\includegraphics[width=1.0\linewidth]{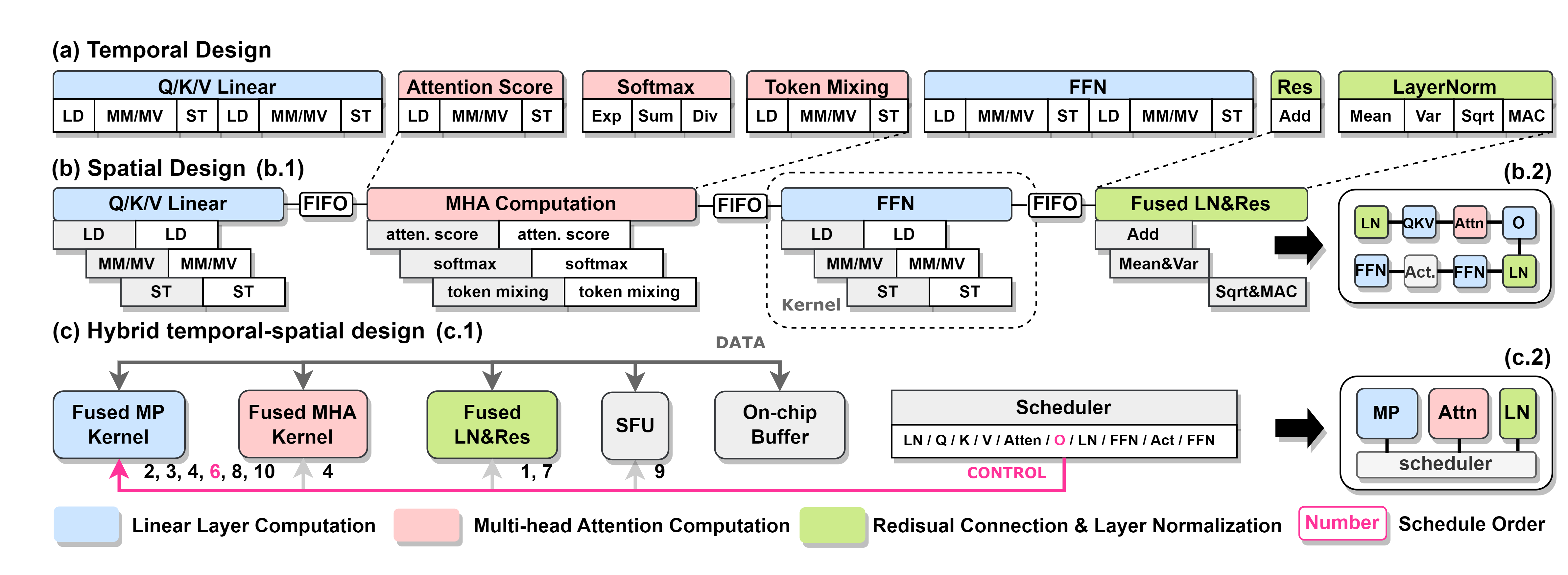}
	% \captionsetup{font=bf}
	\caption{LLM inference under different architectures: (a) execution of temporal architecture. (b) execution of spatial architecture. (c) execution of our proposed hybrid spatial-temporal architecture.}
	\label{fig:mixture_temporal_spatial}
\end{figure*} 

\subsection{Architecture Overview}
\label{sec:architecture_overview} We designed LoopLynx to perform efficient end-to-end speed-up for LLMs by employing a hybrid temporal-spatial design that implements operators as macro dataflow kernels (MDK) (see \secref{sec:Hybriddesign}). \figref{fig:architecture_overview}(a) illustrates the details of our accelerator design, where weights and the KV cache are stored in off-chip high-bandwidth memory (HBM). The accelerator mainly consists of a fused matrix processing (MP) kernel, a fused multi-head attention (MHA) kernel, a fused layernorm and residual (LN\&Res) kernel, and other functional units. These kernels are connected through a shared buffer for data exchange and are managed by a scheduler. Here, "fused" means that several stages are combined to build a dataflow kernel design (see \secref{sec:dataflow}). The router continuously sends and receives neighboring datapacks on the fly.      

\figref{fig:architecture_overview}(b) presents the overall system design. Upon receiving input prompts, the host first embeds each token and then passes it to the accelerator through PCIe for end-to-end inference through $L$ transformer blocks. After completing the prefill stage for the final token, the host synchronizes the model's output and feeds it as input to initiate token generation. For scalability, we integrate a ring network into our dataflow design, with each accelerator node performing symmetrical computations. 

Finally, \figref{fig:architecture_overview}(c) illustrates our model parallelism strategy~\cite{mega}. This strategy distributes the weights of linear layers across devices along the output dimension and employs a head-wise partitioning approach for the KV cache to minimize the memory footprint on each device. For multi-node collaborative inference, the host distributes the same full embedding vector to all nodes, with each node responsible for computing a sub-vector. The synchronization of sub-vectors to reconstruct the full embedding vector can be hidden within computation, thereby minimizing synchronization overhead.

\subsection{Hybrid temporal-spatial design}
\label{sec:Hybriddesign}
In \figref{fig:mixture_temporal_spatial}(a) we demonstrate the inference process of one transformer block in temporal architectures. Take linear layer computations as an example. Model weights are stored in HBM, making the latency of off-chip memory access significant. Temporal architectures use instruction sets
to guide matrix multiplication, involving frequent operations of memory read, compute, and write-back, typically in a serialized manner. This prevents functional units from overlapping during execution, leading to inefficient use of hardware resources and excessive memory access overhead. 

While \figref{fig:mixture_temporal_spatial}(b) shows the spatial architecture, during linear layer computations, matrix multiplication is tiled into blocks, and operations for each block can overlap. Upon completing the current block, functional units can immediately proceed to the next, forming an intra-kernel pipeline. Additionally, the spatial architecture requires the instantiation of all neural network operators to establish an inter-kernel/task-level pipeline (see \figref{fig:mixture_temporal_spatial}(b.1)). This ensures that all kernels run actively at every time slice. However, due to the token-by-token serial decoding pattern in LLMs, connected operators are forced to execute sequentially, restricting dataflow to small, local regions (intra-kernel pipeline) rather than a global scale (inter-kernel/task-level pipeline). This results in poor area utilization and suboptimal latency.

From the two classical architectures mentioned above, we can make the following observations: (1) In temporal architectures, although functional units can be reused, the lack of a dedicated pipeline design leads to serialized execution, resulting in poor peak hardware resource usage and suboptimal performance, as well as excessive memory access overhead. (2) In spatial architectures, if the task-level pipeline cannot be fully established, even though many kernels with the same functionality are instantiated (e.g., the blue-marked kernels for linear layer computation in \figref{fig:mixture_temporal_spatial}(b.2)), they cannot operate simultaneously, also leading to significant resource waste.

To address above limitations, we propose a hybrid spatial-temporal design. Kernels in classical spatial architectures with the same functionality are grouped and implemented as macro dataflow kernels (see \figref{fig:mixture_temporal_spatial}(c.2)), as opposed to the cascaded smaller kernels shown in \figref{fig:mixture_temporal_spatial}(b.2). Dataflow within these MDKs is constructed using the techniques described in \secref{sec:latency_optimizing}. We then employ a scheduler to flexibly organize and reuse these kernels in a temporal manner, achieving much higher peak hardware resource usage during each activation of an MDK compared to classical temporal and spatial architectures. Taking the fused MP kernel as an example, all linear layer computations can be executed using this kernel. At this point, the scheduler enters the $6^{th}$ stage (see \figref{fig:mixture_temporal_spatial}(c.1)) to compute the projection matrix (a linear layer), thus reusing the Fused MP kernel.

\subsection{Latency Optimizing}
\label{sec:latency_optimizing}
\begin{figure}[]
	\centering
	\includegraphics[width=\linewidth]{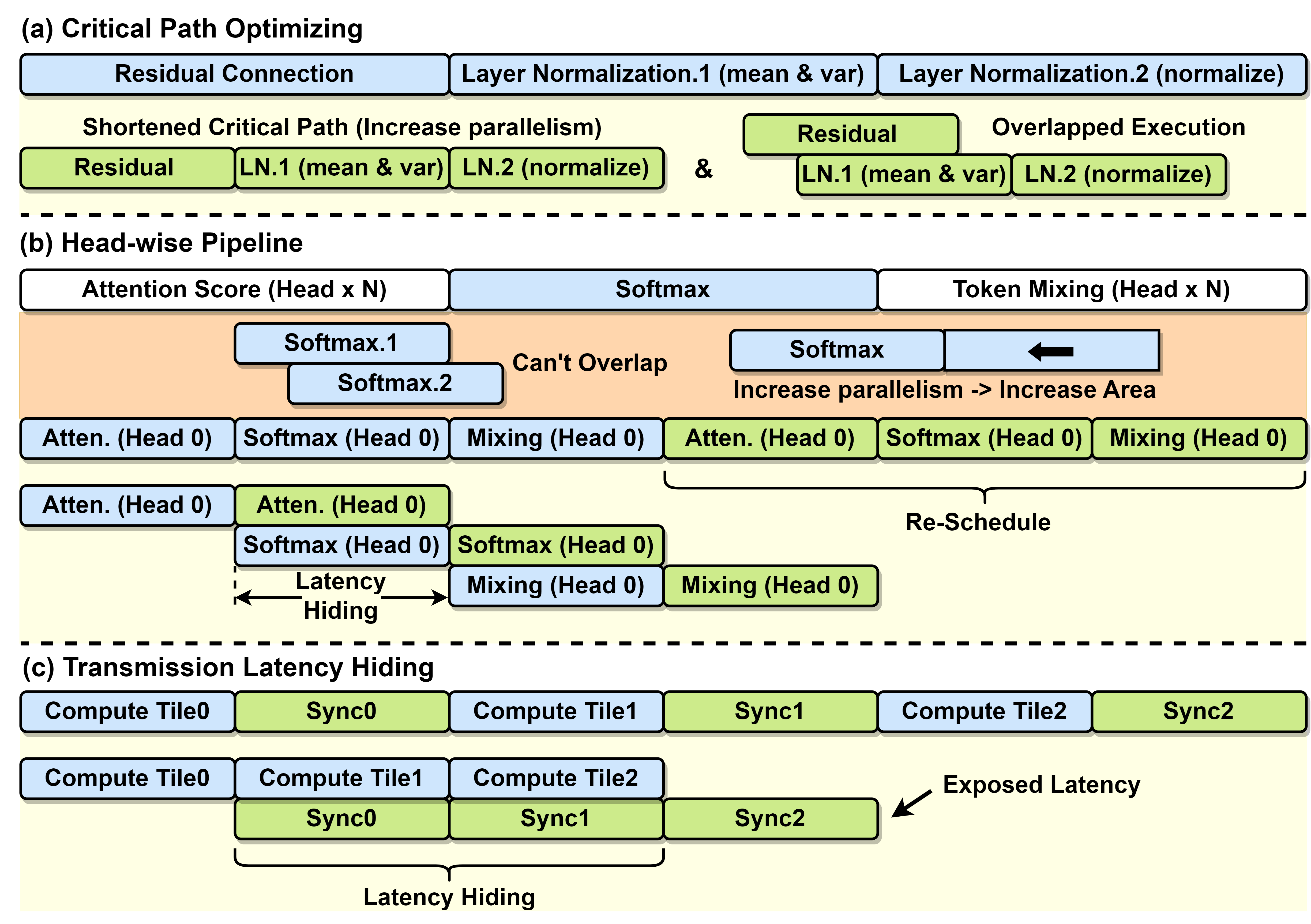}
	% \captionsetup{font=bf}
	\caption{Latency optimization techniques.}
%	\vspace{-1.5\baselineskip}
	\label{fig:latency_optimization}
\end{figure}

\noindent\textbf{Critical Path Optimizing.} We observe that critical path operators—those between each linear layer computation and MHA computation—are as essential as matrix multiplication in determining the overall latency of LLMs. While reducing clock cycles for matrix operations incurs significantly higher resource costs, operators such as residual connections and layer normalization can be parallelized and have their execution overlapped (see \figref{fig:latency_optimization}(a)). This forms a Fused LN\&Res kernel (see Fig. \figref{fig:mixture_temporal_spatial}(c)), achieving improved latency with modest costs.

\noindent\textbf{Head-wise Pipelining.} The calculation of softmax requires obtaining the global sum of exponent values (softmax.1) before generating the weighted score (softmax.2), making it difficult to overlap these two stages to reduce latency (see \figref{fig:latency_optimization}(b)). By reordering the multi-head attention, and forming a head-wise task-level pipeline. The calculation of softmax for $Head_{i-1}$ can be hidden within the attention computation of $Head_i$. This approach avoids the need for excessive hardware resource overhead to reduce the latency of the softmax operator.

\noindent\textbf{Transmission Latency Hiding.} Multi-node synchronization can also employ task splitting and reordering techniques to hide transmission latency. For example, in block matrix multiplications, the data synchronization of the previous block is hidden within the computation of the current block. The actual synchronization overhead occurs only after the completion of the last block matrix (see \figref{fig:latency_optimization}(c)). 

\begin{figure}[]
	\centering
	\includegraphics[width=1.0\linewidth]{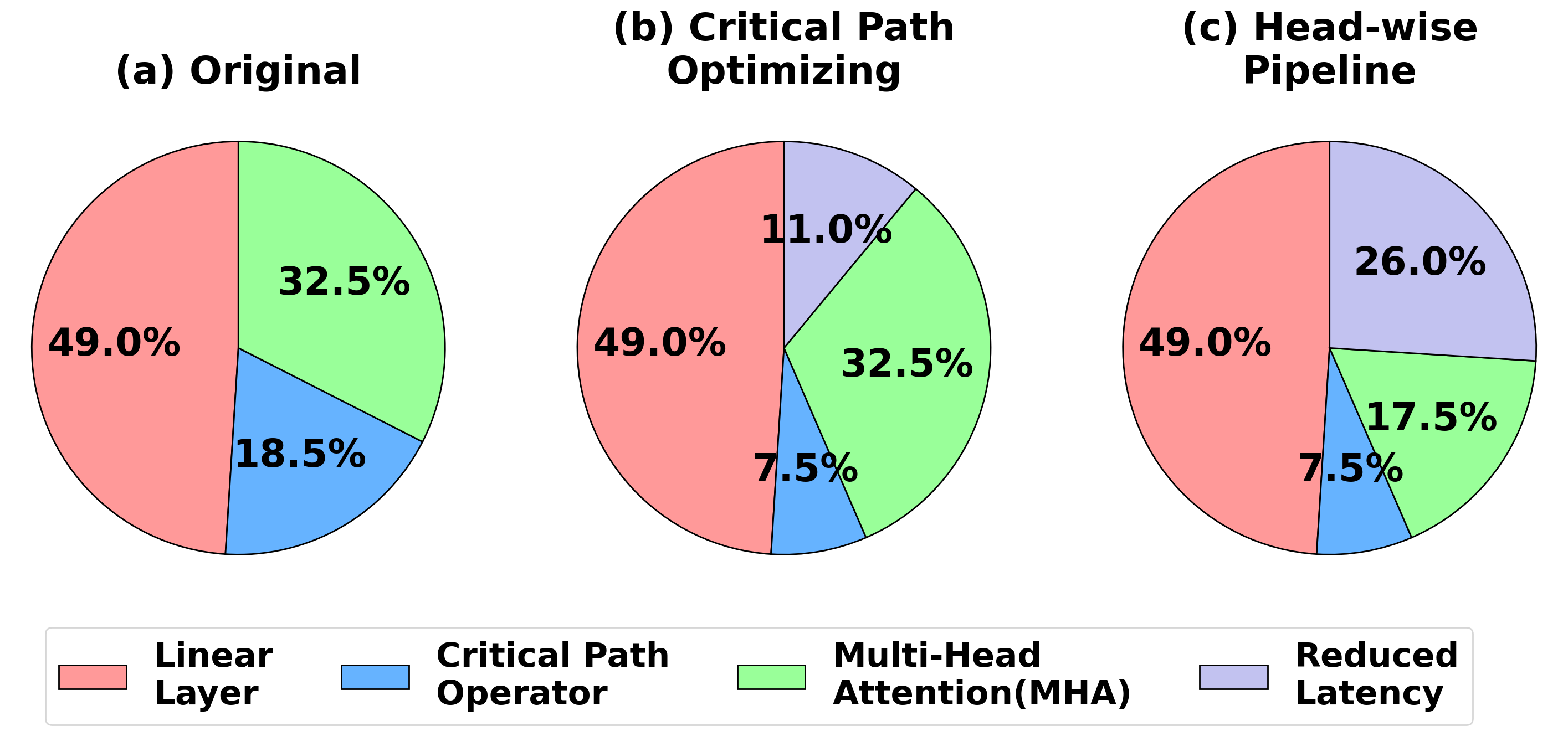}
	% \captionsetup{font=bf}
	\caption{Latency Breakdown of 1-node on GPT-2 and improvement over different optimization techniques.}
	\vspace{-1.2\baselineskip}
	\label{fig:pie}
\end{figure}
\noindent\textbf{Latency Enhancements.} \figref{fig:pie} illustrates the latency breakdown for a single node and the improvements through various optimization techniques presented above. Initially, linear layer and MHA computation accounts for 81.5\% of the latency, while critical path operators contribute 18.5\% (see \figref{fig:pie}(a)). By increasing the parallelism of critical path operators and overlapping the execution of residual connection and layer normalization, we can achieve an 11\% reduction in latency (see \figref{fig:pie}(b)). Further, through the head-wise pipeline (see \figref{fig:pie}(b)), we demonstrate a 15.0\% improvement by effectively hiding the latency of the softmax operator compared to the original version. 

\subsection{Dataflow Architecture}
\label{sec:dataflow}
\begin{figure}[]
	\centering
	\includegraphics[width=0.9\linewidth]{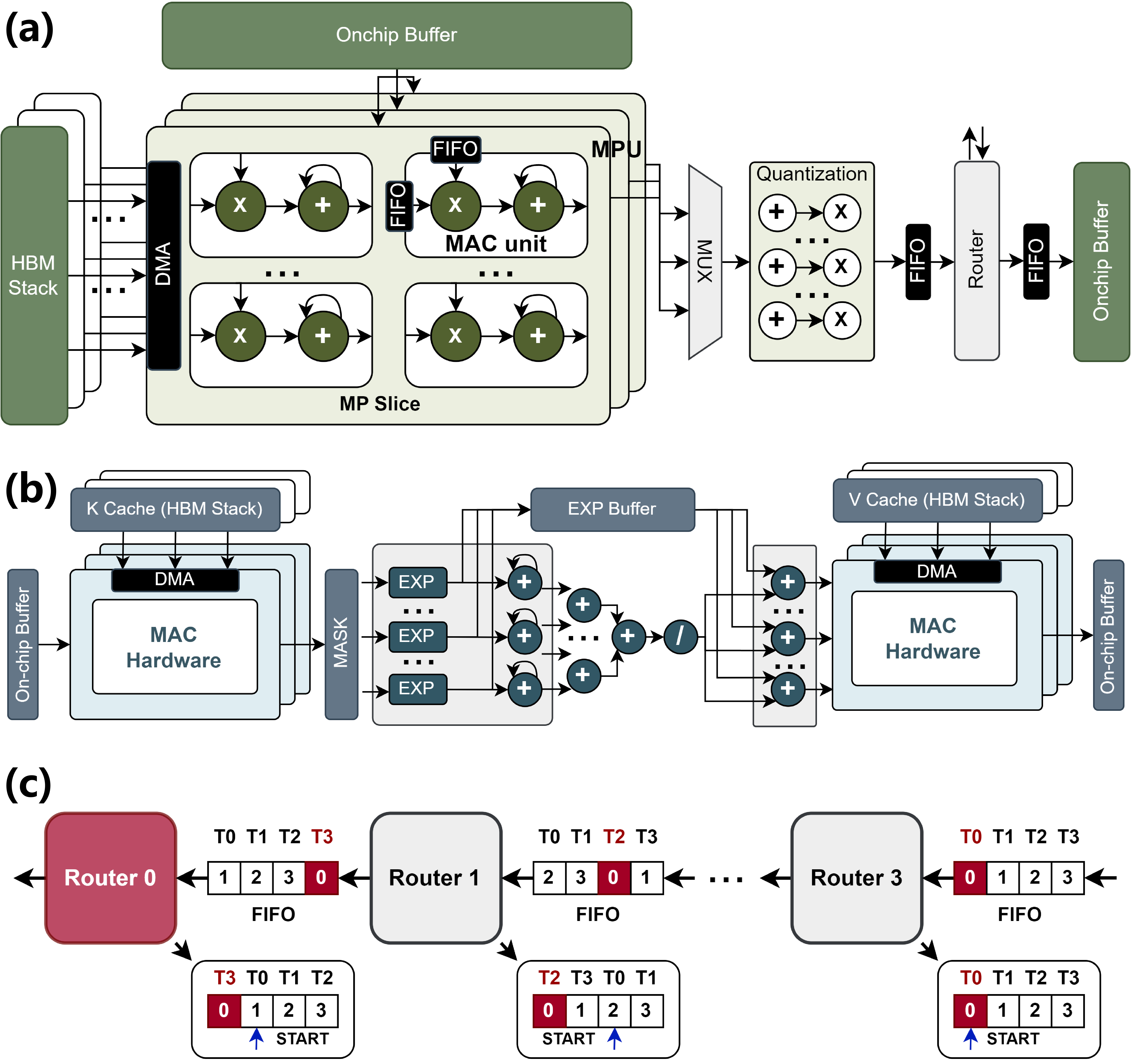}
	% \captionsetup{font=bf}
	\caption{LoopLynx macro dataflow kernels (MDK). (a) Fused Matrix Processing Kernel. (b) Fused Multi-Head Attention Kernel. (c) Router and routing mechanism.}
	\vspace{-1.0\baselineskip}
	\label{fig:details}
\end{figure}

\noindent\textbf{Fused MP Kernel.} As shown in \figref{fig:details}(a), the Fused MP Kernel primarily comprises DMA engines, a matrix processing unit (MPU), a quantization unit, and a router. The MPU performs block matrix-vector multiplication of the tiled weight matrix $W\in \mathbb{Z}^{l_{embed}/n\times l_{embed}}$ and the embedding vector $V\in \mathbb{Z}^{l_{embed}}$. The MPU is designed as accumulator-multiplier based MAC hardware which consists of $n_{channel}$ MP slices, where each one is connected to an HBM channel via the DMA engine. Each MP slice contains $n_{group}$ MAC units. To enhance memory efficiency, the DMA engine runs in burst mode to load concatenated $n_{group}\times$8-bit datapacks onto the chip. We set $n_{group}=32$ to ensure a sufficient burst size. Once MAC units complete $l_{embed}$ MAC operations, the accumulated results are packed and transferred to the quantization unit. Meanwhile, the next block matrix multiplication can proceed. After the quantization unit performs bias addition and quantization, datapacks are forwarded to the router. Benefiting from the decoupled design of the kernel, all above units are connected via FIFOs, thus reducing the place and route (PnR) complexity and enabling the frequency to reach 285 MHz.

\noindent\textbf{Fused MHA kernel.} The Fused MHA kernel consists of two separate MAC hardware implementations, a mask unit and a softmax unit, forming a head-wise task-level pipeline (see \figref{fig:details}(b)). The first MAC hardware is connected to HBM channels used as key cache and computes attention scores for each head while passing these scores to the mask unit. The mask unit ensures that only forward attention is kept during inference, and then passes the result to the softmax unit to compute weighted attention scores. After that, the output is sent to the second MAC hardware, where cached values are loaded to perform token mixing.

\noindent\textbf{Routing mechanism.} The router operates in simplex mode. As illustrated in \figref{fig:details}(c), with four nodes, the process involves four rounds of buffer writing followed by reading. During each round, each node writes $n$ datapacks to its successor node and reads $n$ datapacks from its predecessor node. This ensures that each node synchronizes the datapacks from the other nodes. Meanwhile, each router maintains an offset based on the node ID, and the router continuously writes the received datapacks into the buffer starting from this offset. This ensures that all buffers maintain consistent data after four rounds of synchronization.

\begin{figure}[]
	\centering
	\begin{minipage}{0.3\linewidth}
			\centering
			\includegraphics[width=\linewidth]{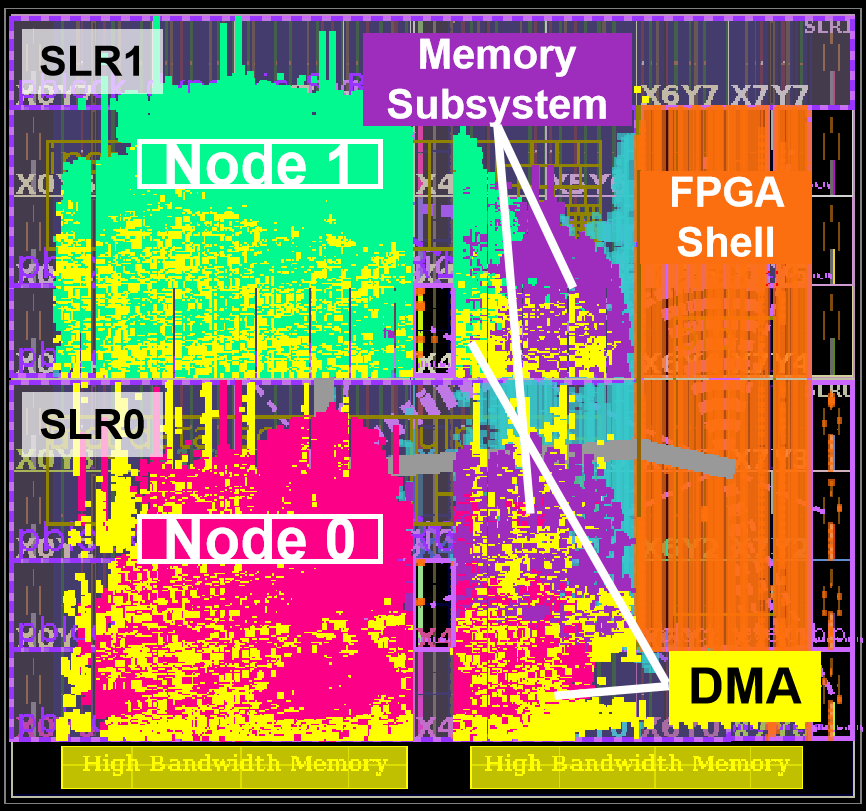}
		\end{minipage}\hfill
	\begin{minipage}{0.67\linewidth}
			\centering
			\resizebox{\linewidth}{!}{%
					\begin{tabular}{|l|c|c|c|c|}
							\hline
							Component & DSP & LUT & FF & BRAM  \\\hline
							Fused MP Kernel & 522 & 34K & 56K & 241 \\\hline
							Fused MHA Kernel & 382 & 38K & 45K & 16  \\\hline
							Fused LN Kernel & 192 & 23K & 30K & 240  \\\hline
							DMA & 0 & 16K & 28K & 97 \\\hline
							Other Kernels/Buffer & 32 & 17K & 26K & 1 \\\hline
							Device Total & 1132 & 312K & 478K & 924.5  \\\hline
							Accelerator Total & 1128 & 128K & 185K & 595  \\\hline
					\end{tabular}}
		\end{minipage}
	% \captionsetup{font=bf}
	\caption{FPGA layout of dual-node setting and resource utilization on Xilinx Alveo U50.}
	\label{fig:layout}
\end{figure}

\begin{table}[]
	\centering
	% \captionsetup{font=bf}
	\caption{Comparison of GPU and FPGA platforms.\protect\footnotemark[1]}
	\resizebox{\columnwidth}{!}{
		\begin{tabular}{lcccccc}
			\toprule
			\textbf{Platform} & \textbf{Process} & \textbf{Frequency} & \textbf{Computing Units} & \textbf{Bandwidth} & \textbf{TDP}\\
			\midrule
			Nvidia A100 & 7nm & 1065MHz & 432 Tensor Cores & 1935 GB/s & 300W \\
			Xilinx Alveo U280 & 16nm & 200-300MHz & 9024 DSPs & 460 GB/s & 215W \\
			Xilinx Alveo U50 & 16nm & 200-300MHz & 5952 DSPs & 201 GB/s & 75W \\
			\bottomrule
		\end{tabular}
	}
	\label{tab:comparison}
\end{table}

\begin{table*}[!t]
	\centering
	% \captionsetup{font=bf}
	\caption{Comparison of FPGA Implementations.}
	\renewcommand{\arraystretch}{0.8} 
	\resizebox{\textwidth}{!}{%
		\begin{tabular}{@{}cccccccccc@{}}
			\toprule
			Architecture                      & \# Nodes         & Freq.                     & Quantization          & Token Latency & DSP & BRAM & LUT & FF & URAM \\ \midrule
			\multirow{3}{*}{LoopLynx} & 4 Nodes (U50 x2) & \multirow{3}{*}{285 MHz} & \multirow{3}{*}{W8A8} & 2.55 ms                & 2264   & 1609     & 624K   & 954K  & 8    \\
			& 2 Nodes (U50 x1) &         &         & 3.85 ms & 1132 & 924.5 & 312K & 478K  & 4   \\
			& 1 Node (U50 x1) &         &         & 6.59 ms & 568 & 641  & 220K & 313K  & 4   \\ \midrule
			Temporal Architecture~\cite{dfx}         & U280            & 200 MHz & Float16 & 5.37 ms & 3533 & 1192  & 520K & 1107K & 104 \\ \midrule
			Spatial Architecture~\cite{Spatial}& U280            & 245 MHz & W8A8    & 4.17ms  & 1780 & 389   & 653K & 569K  & 111 \\ \bottomrule
		\end{tabular}%
	}
	\label{tab:compare}
\end{table*}
\subsection{Evaluation Setup}
To validate the feasibility of LoopLynx architecture, we implement all the modules of LoopLynx in C++ with Vitis HLS and synthesize the accelerator with Vivado on AMD Alveo U50 FPGA~\cite{amd_alveo_u50}. To demonstrate the scalability, we scale LoopLynx architecture into multiple accelerator nodes which are distributed and interconnected across multiple FPGAs using AXI-Stream for ring connections (see  \ref{sec:architecture_overview}). One Alveo U50 FPGA is composed of two super logic regions (SLRs). According to our implementation as shown in \figref{fig:layout}, one accelerator node can fit within one SLR region of the Alveo U50 FPGA. Therefore, we deploy two accelerator nodes across two SLRs in one Alveo U50 FPGA. We measured inference latency using cycle-accurate simulation, fully accounting for the per-channel HBM bandwidth (peak 8.49 GB/s) and network bandwidth (peak 8.49 GB/s), to explore the potential of implementing large-scale accelerator nodes across multiple FPGAs\protect\footnotemark[2].

We select GPT-2 (345M) scale model to evaluate the inference speed, both our accelerator and GPU application use the smoothquant~\cite{smoothquant} W8A8 quantization scheme. We chose A100 as our GPU baseline. Since PyTorch employs pseudo-int8 quantization, we used the torch-int~\cite{smoothquant} library to fully leverage GPU performance during inference. For the FPGA baseline, we compared ours with the state-of-the-art temporal architecture DFX~\cite{dfx} and spatial architecture~\cite{Spatial}. The implementation in \cite{Spatial} has separate versions for prefill and decode, so we calculated a weighted per-token processing latency. The author reports that the resource utilization of these two implementations is similar. For DFX, we compared it with their single U280 implementation. \tabref{tab:comparison} shows the difference in compute ability and memory bandwidth between hardware.

\footnotetext[1]{ \url{https://adaptivesupport.amd.com/s/article/75222}}
\footnotetext[2]{ \url{https://github.com/zjnyly/LoopLynx}}

\subsection{Evaluation Results}
\label{sec:scalability}
\noindent\textbf{Comparison with FPGAs.} First, we compare our scalable hybrid spatial-temporal design LoopLynx with the temporal architecture DFX~\cite{dfx} and the spatial architecture~\cite{Spatial}. \tabref{tab:compare} provides a detailed comparison of our scaled multi-node design against temporal and spatial architectures~\cite{dfx,Spatial} in terms of average per-token latency and resource utilization. 
From \tabref{tab:compare}, we can make the following observations: 
(1) Our scalable architectures with both 2-node and 4-node implementations achieve significant speed-up compared to temporal architecture DFX~\cite{dfx} and spatial architecture~\cite{Spatial}, while maintaining modest FPGA resource usage. Specifically, our 2-node implementation brings 1.39x and 1.08x faster speed-up when compared to temporal architecture DFX~\cite{dfx} and spatial architecture~\cite{Spatial}, respectively. For 4-node implementation, LoopLynx can respectively gain 2.11x and 1.64x acceleration over two FPGA accelerator baselines~\cite{dfx,Spatial} while maintaining reasonable resource utilization. Such acceleration benefits from
not only the decoupled dataflow architecture design (see \secref{sec:dataflow}) which can achieve high clock speed
 but also multi-FPGA distributed architecture design (see \secref{sec:architecture_overview}) which can fully utilize the distributed accelerators for massive kernel parallelism.
(2) For single-node implementation, LoopLynx is slightly slower than two FPGA accelerator baselines~\cite{dfx,Spatial}. However, LoopLynx is far more resource-efficient than two FPGA accelerator baselines, offering a viable option for cost-efficient edge devices with reasonable token latency. 

\noindent\textbf{Comparison with GPU.}
\begin{figure*}[h!]
	\centering
	\includegraphics[width=0.9\linewidth]{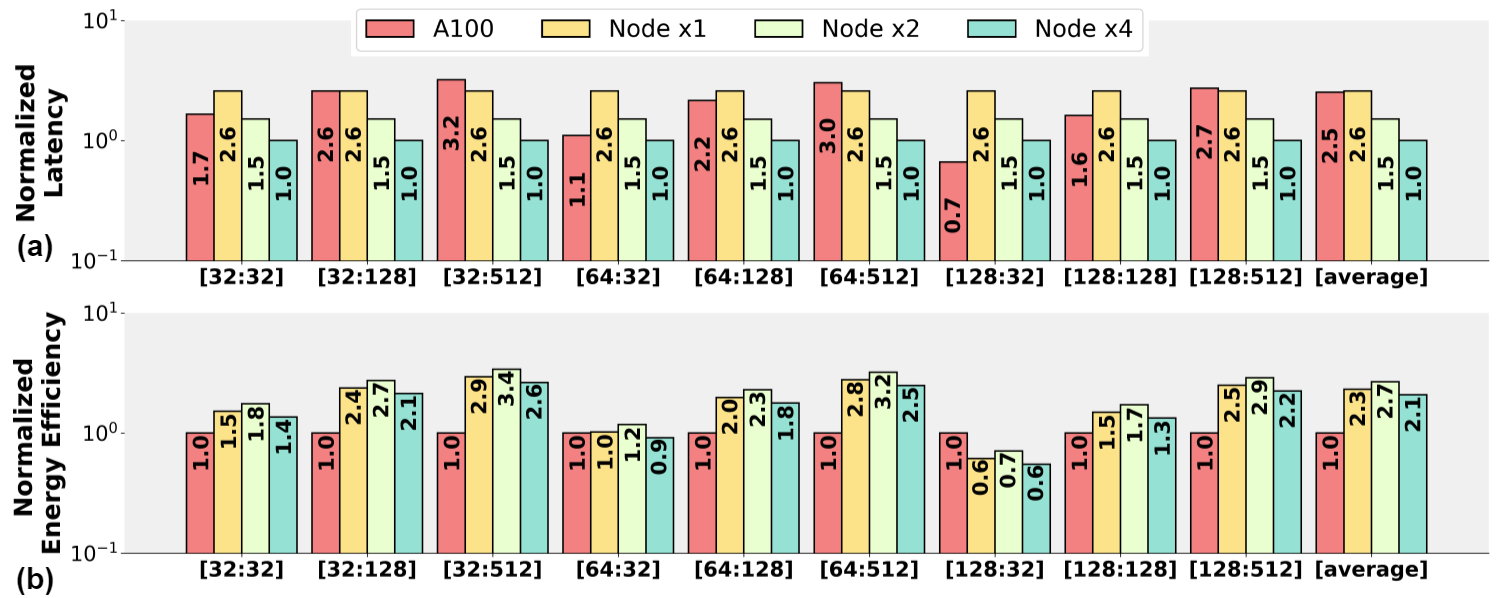}
	% \captionsetup{font=bf}
	\caption{Comparison of (a) normalized inference latency and (b) normalized energy efficiency between LoopLynx and Nvidia A100 GPU across various [prefill size: decode size] sequences, using logarithmic scale on the y-axis.}
	% \vspace{-1.0\baselineskip}
	\label{fig:latency}
\end{figure*}
Next, we compare the latency and energy efficiency by running GPT-2 on LoopLynx accelerators and Nvidia A100 with various [input:output] length settings. We use Nvidia built-in function in \cite{NVIDIA_SMI} and Xilinx power analysis tool to obtain the power consumption for two implementations. \figref{fig:latency} shows the comparison results. In \figref{fig:latency}, overall latency (ms) is normalized to 4-node implementations where higher values indicate slower speed. The energy efficiency (token/J) is normalized to GPU implementations, where higher values indicate better efficiency.
From \figref{fig:latency}(a), we find that our scalable LoopLynx accelerators with 2-node and 4-node implementations show great advantages compared with GPU implementations in scenarios like code generation and chatbots which require long text generation (on the settings of [32:512], [64:512], [128:512]). On average, LoopLynx accelerators with 2-node and 4-node implementations can achieve 1.67x and 2.52x speed-up compared to A100, respectively. For the setting of [128:32], A100 performs better over LoopLynx accelerator. This is due to the fact that GPUs are more powerful in batched processing during the prefill stage. 
From \figref{fig:latency}(b),  our LoopLynx accelerator also demonstrates better energy efficiency. On average, LoopLynx accelerator with three distributed implementations can achieve 2.3x, 2.7x and 2.1x when compared to A100, respectively. Notably, LoopLynx accelerator with 2-node implementation maintains the highest energy efficiency among other settings, striking a balance between latency and resource utilization. 
Thus, different node settings offer various design options between processing speed and energy efficiency, highlighting the potential of FPGAs for LLM serving.

\begin{table}[]
	\centering
	% \captionsetup{font=bf}
	\caption{Throughput and Scalability.}
	\renewcommand{\arraystretch}{0.8} % 调整行高
	\begin{tabular}{*{3}{>{\centering\arraybackslash}p{\dimexpr \linewidth/3-2\tabcolsep\relax}}}
		\toprule % 顶部粗线
		\textbf{\# Nodes} & \textbf{Tokens Per Second} & \textbf{Speed-up\protect\textsuperscript{1}} \\
		\midrule % 中间细线
		1-node & 151.7 token/s & - \\
		2-node & 259.7 token/s & 1.71x \\
		4-node & 392.2 token/s & 1.51x \\
		\bottomrule % 底部粗线
	\end{tabular}
	\vspace{-1.5\baselineskip}
	\label{tab:scalablity}
\end{table}

\noindent\textbf{Scalability Analysis.} We further discuss the scalability and interconnect overhead of LoopLynx. 
\tabref{tab:scalablity} shows the throughput of deploying the GPT-2 model on our scalable LoopLynx accelerators. From \tabref{tab:scalablity}, we can see that 2-node and 4-node implementations can achieve 1.71x and 1.51x speed-ups. The speed-up factor does not present linear growth (close to 2x), primarily due to two reasons: (1) Operators on the critical path cannot be distributed across multiple devices for cooperative computing. (2) When distributing matrix multiplication tasks to 2–4 nodes, the overlapping tasks require fewer than 1 node, exposing the latency of quantization and synchronization. This can be mitigated by increasing the workload assigned to each node.

\footnotetext[1]{Speed-up on 2-node implementation is calculated with respect to 1-node implementation, while speed-up on 4-node implementation is relative to 2-node implementation.}

\section{Conclusion}
This paper introduces LoopLynx, a scalable dataflow architecture designed for efficient LLM inference. We propose a hybrid spatial-temporal design to capitalize on dataflow's throughput advantages, thereby enhancing end-to-end inference performance under the serialized LLM decoding pattern. With our scalable design,  LoopLynx can be expanded to multi-FPGAs. Our dual-FPGA configuration achieves an average 2.52x speed-up across diverse usage scenarios compared to the Nvidia A100, while consuming  48.1\% of its energy, revealing the potential of FPGAs for LLM serving.

\section*{Acknowledgment}
This research was supported by the National Natural Science Foundation of China under Grant
92470202, the Fund of National Key Laboratory of Multispectral Information Intelligent Processing
Technology (No. 202410487201).

{
	\footnotesize
	%\scriptsize
	%\small
	\bibliographystyle{unsrt}
	%\bibliographystyle{IEEEtran}
	% \normalem
	
	\bibliography{./ref}
}

\end{document}